\documentclass[aps,prl,showpacs,twocolumn]{revtex4}
\usepackage{amsmath}
\usepackage{graphicx}
\usepackage[ansinew]{inputenc}
\usepackage{array}
\usepackage{color}
\usepackage{amsxtra}
\usepackage{amstext}
\usepackage{amssymb}
\usepackage{latexsym}
\usepackage{dsfont}
\usepackage{bbold}
\usepackage{isomath}
\usepackage{pzccal}
\begin{document}

\title{Ground-state OH molecule in combined electric and magnetic fields:
Analytic solution of the effective Hamiltonian}

\author{M. Bhattacharya and Z. Howard}
\affiliation{School of Physics and Astronomy, Rochester Institute of Technology, 84 Lomb Memorial Drive,
Rochester, NY 14623, USA}
\author{M. Kleinert}
\affiliation{Department of Physics, Willamette University, 900 State Street, Salem, OR 97301, USA}

\date{\today}
\begin{abstract}
The OH molecule is currently of great interest from the perspective of ultracold chemistry, quantum fluids,
precision measurement and quantum computation. Crucial to these applications are the slowing, guiding,
confinement and state control of OH, using electric and magnetic fields. In this article, we show that the
corresponding eight-dimensional effective ground state Stark-Zeeman Hamiltonian is exactly solvable and
explicitly identify the underlying chiral symmetry. Our analytical solution opens the way to
insightful characterization of the magnetoelectrostatic manipulation of ground state OH. Based on our
results, we also discuss a possible application to the quantum simulation of an imbalanced Ising magnet.
\end{abstract}

\maketitle
The OH molecule in its ground $X^{2}\mathrm{\Pi}_{3/2}$ state is presently widely employed in investigations of
ultracold chemistry \cite{Ticknor2005,Avdeenkov2003,Sawyer2008,Tscherbul2010}, precision measurements
\cite{Hudson2006,Kozlov2009}, and quantum computation \cite{Lev2006}. Particularly interesting is the recently
implemented evaporative cooling of OH close to Bose-Einstein condensation \cite{Quemener2012}.
With such experiments underway, the exploration of quantum degeneracy and molecular optics \cite{Meijer2009}
with OH should shortly become reality.

A substantial reason behind the suitability of OH as a workhorse for these experiments
is the fact that it is a polar paramagnetic molecule, i.e. it carries both electric and magnetic dipole moments.
Electric and magnetic fields can therefore be used to slow, guide, confine and generally
manipulate OH \cite{Bochinski2004,Meerakker2005,Sawyer2007,Stuhl2012,Schmelcher2013}. It follows that a
quantitative as well as qualitative understanding of the corresponding Stark-Zeeman spectrum is of great relevance.

In this article we present the exact solution of the eight-dimensional Stark-Zeeman Hamiltonian of OH in its
$X^{2}\mathrm{\Pi}_{3/2}$ ground state \cite{Stuhl2012} and identify the intriguing underlying symmetry.
This molecular Hamiltonian is an effective one, neglects hyperfine structure, and, has been used to numerically model
experimental data accurately \cite{Sawyer2007,Stuhl2012,Quemener2012}. However, there is interest in analytic solutions
also: during the preparation of this article, the field-dependent part of the Hamiltonian was
diagonalized exactly in an insightful article by Bohn and Quemener \cite{Bohn2013}.

Based on our analysis, we suggest that the OH molecule may be used to simulate a mixed spin Ising magnet, which is
of interest in condensed matter physics \cite{Buendia1999}. Another use for our results is a realistic theory of
nonadiabatic processes in traps, which so far has relied on a simplified four-dimensional model of the
OH ground state \cite{Lara2008}. Our work may also be of relevance to atmospheric \cite{Wayne1990}, interstellar
\cite{Wardle2002} and combustion physics \cite{Smith2002}, where OH plays an important
role. Lastly, we hope that our results will usefully add to the handful of exact solutions available
for molecules, especially in strong fields \cite{SchmelcherBook}.

We begin with the Stark-Zeeman Hamiltonian for OH in the $X^{2}\mathrm{\Pi}_{3/2}$ state, as presented earlier
\cite{Stuhl2012}
\begin{equation}
\label{eq:H1}
H=H_{o}-\vec{\mu}_{e}\cdot \vec{E}-\vec{\mu}_{b}\cdot \vec{B},
\end{equation}
where $H_{o}$ is the field-free Hamiltonian,
$\vec{\mu}_{e}$ and $\vec{\mu}_{b}$ are the electric and magnetic dipole moments of the molecule, respectively, and
$\vec{E}$ $[\vec{B}]$ is the electric [magnetic] field imposed on the molecule. This model is
valid when hyperfine structure is negligible, such as for electric fields stronger than 1kV/cm and magnetic fields above
100G \cite{Ticknor2005}, or for OH vapor temperatures higher than 5mK. A number of experiments lie in these regimes
\cite{Meerakker2005,Sawyer2008,Stuhl2012,Quemener2012}.

The matrix representation of the Hamiltonian in Eq.~(\ref{eq:H1}) can be obtained using the Hund's
case (a) parity basis $|J,M,\bar{\Omega},\epsilon \rangle$ suggested by Lara \textit{et al.}, where $J=3/2$ is the total
angular momentum, $M$ its projection in the laboratory frame, $\bar{\Omega}$ its projection on the internuclear
axis, and $\epsilon=\{e,f\}$ is the $e-f$ symmetry \cite{Lara2008}. Following Ref. [\onlinecite{Lara2008}], both the electric
and magnetic moments are assumed to lie along the axis of the molecule, and the magnetic field is chosen along the
laboratory $z$ axis, with which the electric field makes an angle $\theta$. With these assumptions, the Hamiltonian matrix
has been found to be \cite{Stuhl2012}
\begin{widetext}
\begin{equation}
\label{eq:Hmatrix} H_{M} = \\
\small
\begin{pmatrix}
  -\frac{\hbar\Delta}{2}-\frac{6}{5}\mu_{B}B      &     0         &    0        &       0 & \frac{3}{5}\mu_{e}E\cos\theta   & -\frac{\sqrt{3}}{5}\mu_{e}E\sin\theta            &    0         & 0  \\
           0     & -\frac{\hbar\Delta}{2}-\frac{2}{5}\mu_{B}B    & 0    &      0 & -\frac{\sqrt{3}}{5}\mu_{e}E\sin\theta  & \frac{1}{5}\mu_{e}E\cos\theta            &  -\frac{2}{5}\mu_{e}E\sin\theta           &  0 \\
           0     & 0       &  -\frac{\hbar\Delta}{2}+\frac{2}{5}\mu_{B}B   &   0  &  0      &   -\frac{2}{5}\mu_{e}E\sin\theta            &             -\frac{1}{5}\mu_{e}E\cos\theta       & -\frac{\sqrt{3}}{5}\mu_{e}E\sin\theta  \\
           0&0&0&-\frac{\hbar\Delta}{2}+\frac{6}{5}\mu_{B}B &0 &0 &-\frac{\sqrt{3}}{5}\mu_{e}E\sin\theta &-\frac{3}{5}\mu_{e}E\cos\theta \\
\frac{3}{5}\mu_{e}E\cos\theta  & -\frac{\sqrt{3}}{5}\mu_{e}E\sin\theta  &       0        & 0  & \frac{\hbar\Delta}{2}-\frac{6}{5}\mu_{B}B  &             0&     0        &  0 \\
-\frac{\sqrt{3}}{5}\mu_{e}E\sin\theta   & \frac{1}{5}\mu_{e}E\cos\theta & -\frac{2}{5}\mu_{e}E\sin\theta & 0 & 0  & \frac{\hbar\Delta}{2}-\frac{2}{5}\mu_{B}B             &   0          &  0 \\
       0    &  -\frac{2}{5}\mu_{e}E\sin\theta & -\frac{1}{5}\mu_{e}E\cos\theta  & -\frac{\sqrt{3}}{5}\mu_{e}E\sin\theta & 0 & 0   &\frac{\hbar\Delta}{2}+\frac{2}{5}\mu_{B}B   & 0 \\
       0    &  0     & -\frac{\sqrt{3}}{5}\mu_{e}E\sin\theta  & -\frac{3}{5}\mu_{e}E\cos\theta &   0  & 0 &0 &\frac{\hbar\Delta}{2}+\frac{6}{5}\mu_{B}B   \\
\end{pmatrix},
\end{equation}
\end{widetext}
where $\Delta$ is the lambda-doubling parameter, $\mu_{B}$ is the Bohr magneton, $\mu_{e}$ the molecular electric dipole moment,
and $E$ [$B$] are the magnitudes of the electric [magnetic] fields.

\begin{figure}[]
\includegraphics[width=0.4\textwidth]{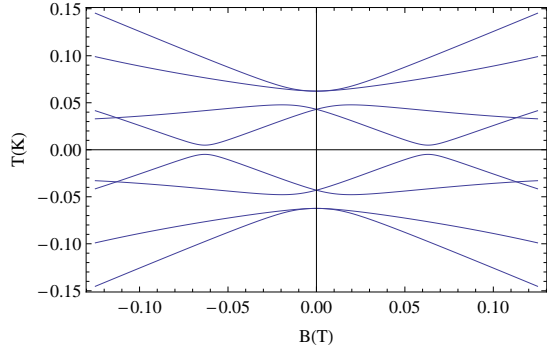}
\caption{Numerical plot of the eigenvalues of Eq.~(\ref{eq:Hmatrix}) with $\Delta=2\pi\times1.667$GHz, $\mu_{e}=1.66$D,
$E=2$kV/cm, and $\theta=\pi/2$ \cite{Stuhl2012}. The horizontal axis denotes the magnetic field in Tesla and the vertical axis
corresponds to energy in Kelvin.}
\label{fig:P1}
\end{figure}

To the best of our knowledge, the matrix $H_{M}$ of Eq.(\ref{eq:Hmatrix}) has only been diagonalized numerically
so far \cite{Sawyer2007,Stuhl2012,Quemener2012,Bohn2013}. This is not surprising, as in general one may not expect an $8\times 8$
matrix to yield analytic eigenvalues. However, a numerical plot of the spectrum, shown in Fig.~\ref{fig:P1},
presents a tantalizing symmetry about the zero energy (horizontal) axis: if $\lambda$ is an eigenvalue, so is $-\lambda.$
This suggests that the characteristic polynomial $P(\lambda)$ of $H_{M}$,
\vspace{-0.1in}
\begin{equation}
\label{eq:poly}
P(\lambda)=|H_{M}-\lambda I|=\displaystyle\sum_{n=0}^{8}p_{n}\lambda^{n}.
\end{equation}
might be even in $\lambda$. Indeed, a straightforward calculation shows that the coefficients of all the odd powers of
$\lambda$ vanish:
\begin{equation}
p_{1}=p_{3}=p_{5}=p_{7}=0.
\end{equation}
Therefore, $P(\lambda)$, which is an octic in $\lambda$, can be written as a quartic in $\lambda^{2}$,
\begin{equation}
\label{eq:octic}
P(\lambda)=p_{0}+p_{2}\lambda^{2}+p_{4}(\lambda^{2})^{2}+p_{6}(\lambda^{2})^{3}+(\lambda^{2})^{4},
\end{equation}
and thus its roots can be found analytically \cite{Merriman1892}. The even coefficients are
\vspace{-0.08in}
\begin{eqnarray}
\label{eq:peven}
p_{0}&=&\frac{1}{10^{8}}\left\{81 \tilde{B}^{8}+324 \tilde{B}^{4}\tilde{E}^{4}\right.
+81 \tilde{E}^{8}-180 \tilde{B}^{6}\tilde{\Delta}^{2}\nonumber\\
&&+756\tilde{B}^{4}\tilde{E}^2\tilde{\Delta}^{2}-756\tilde{B}^{2}\tilde{E}^{4}\tilde{\Delta}^{2}
+180\tilde{E}^{6}\tilde{\Delta}^{2}\nonumber\\
&&+118\tilde{B}^{4}\tilde{\Delta}^{4}-264\tilde{B}^{2}\tilde{E}^{2}\tilde{\Delta}^{4}+118\tilde{E}^{4}\tilde{\Delta}^{4}\nonumber\\
&&-20\tilde{B}^{2}\tilde{\Delta}^{6}+20\tilde{E}^{2}\tilde{\Delta}^{6}+\tilde{\Delta}^{8}
\left.-4\tilde{B}^{2}\tilde{E}^{2}\left(81\tilde{B}^{4}\right.\right.\nonumber\\
&&\left.+81\tilde{E}^{4}+54\tilde{B}^{2}\tilde{\Delta}^{2}-54\tilde{E}^{2}\tilde{\Delta}^{2}-7\tilde{\Delta}^{4}\right)\cos 2\theta\nonumber\\
&&\left.+162\tilde{B}^{4}\tilde{E}^{4}\cos 4\theta\right\},\nonumber\\
p_{2}&=&\frac{1}{50000}\left\{-9\tilde{B}^{6}-9\tilde{E}^{6}-\frac{1}{5}\tilde{\Delta}^{6}-\frac{59}{5}\tilde{E}^{4}\tilde{\Delta}^{2}\right.\nonumber\\
&&-3\tilde{E}^{2}\tilde{\Delta}^{4}-9\tilde{B}^{4}\tilde{E}^{2}-\frac{23}{5}\tilde{B}^{4}\tilde{\Delta}^{2}-9\tilde{B}^{2}\tilde{E}^{4}\nonumber \\
&&+\tilde{B}^{2}\tilde{\Delta}^{4}+\frac{48}{5}\tilde{B}^{2}\tilde{E}^{2}\tilde{\Delta}^{2}\nonumber\\
&&+2\tilde{B}^{2}\tilde{E}^{2}\left.\left(9\tilde{B}^{2}+9\tilde{E}^{2}+\frac{17}{5}\tilde{\Delta}^{2}\right)\cos 2\theta \right\},\\
p_{4}&=&\frac{1}{5000}\left\{59\tilde{B}^{4}+36\tilde{B}^{2}\tilde{E}^{2}+10\tilde{B}^{2}\tilde{\Delta}^{2}\right.\nonumber\\
&&\left.-82\tilde{B}^{2}\tilde{E}^{2}\cos 2\theta+59\tilde{E}^{4}+30 \tilde{E}^{2}\tilde{\Delta}^{2}+3\tilde{\Delta}^{4}\right\},\nonumber\\
p_{6}&=&-\frac{1}{5}\left\{\tilde{B}^{2}+\tilde{E}^{2}+\frac{1}{5}\tilde{\Delta}^{2}\right\},\\
\nonumber
\end{eqnarray}
where $\tilde{B}=4\mu_{B}B$, $\tilde{E}=2\mu_{e}E$, and $\tilde{\Delta}=5\hbar \Delta$. The eigenvalues are
\vspace{-0.01in}
\begin{eqnarray}
E_{3/2,f}&=&\sqrt{-\frac{p_{6}}{4} + \frac{\sqrt{g_{1}}}{2}+ \frac{\sqrt{g_{2}+g_{3}}}{2}}=-E_{-3/2,e},\\
E_{3/2,e}&=&\sqrt{-\frac{p_{6}}{4} + \frac{\sqrt{g_{1}}}{2}- \frac{\sqrt{g_{2}+g_{3}}}{2}}=-E_{-3/2,f},\\
E_{1/2,f}&=&\sqrt{-\frac{p_{6}}{4} - \frac{\sqrt{g_{1}}}{2}+ \frac{\sqrt{g_{2}-g_{3}}}{2}}=-E_{-1/2,e},\\
E_{1/2,e}&=&\sqrt{-\frac{p_{6}}{4} - \frac{\sqrt{g_{1}}}{2}- \frac{\sqrt{g_{2}-g_{3}}}{2}}=E_{-1/2,f},\\
g_1&=&-\frac{2p_{4}}{3}+\frac{p_{6}^{2}}{4}+\frac{2^{1/3}h_{2}}{3h_{3}}+\frac{h_{3}}{2^{1/3}3},\nonumber\\
g_2&=&-\frac{4p_{4}}{3}+\frac{p_{6}^{2}}{2}-\frac{2^{1/3}h_{2}}{3h_{3}}-\frac{h_{3}}{2^{1/3}3},\nonumber\\
g_3&=&\frac{-8 p_2 + 4 p_4 p_6 - p_{6}^{3}}{4\sqrt{g_{1}}},\nonumber\\
h_1&=&27 p_2^2 - 72 p_0 p_4 + 2 p_4^3 -9 p_2 p_4 p_6 + 27 p_0 p_6^2,\nonumber\\
h_2&=&12 p_0 + p_4^2 - 3 p_2 p_6,\nonumber\\
h_3&=&\left(h_1+\sqrt{h_1^2-4 h_2^3}\right)^{1/3}.\\
\nonumber
\end{eqnarray}
We have verified that the analytical eigenvalues reproduce exactly the numerical spectrum of Fig.~\ref{fig:P1},
and of other references \cite{Stuhl2012,Quemener2012}.

We now proceed to formally investigate the source of
the reflection symmetry of the spectrum, which is responsible for making the problem exactly solvable. In order
to do this it is crucial to note that that the Hamiltonian of Eq.~(\ref{eq:Hmatrix}) can be written as
\begin{eqnarray}
\label{eq:HMDP1}
H_{M}&=&\frac{\hbar\Delta}{2}\left(
                             \begin{array}{cc}
                               -I_{4} & 0 \\
                               0 & I_{4} \\
                             \end{array}
                           \right)
                           +\frac{4\mu_{B}B}{5\hbar}\left(
                                                   \begin{array}{cc}
                                                     -J_{z} & 0 \\
                                                     0 & -J_{z} \\
                                                   \end{array}
                                                 \right)\\
  &&+\frac{2\mu_{e}E}{5\hbar}\left(
                             \begin{array}{cc}
                               0 & \cos\theta J_{z}-\sin\theta J_{x}\nonumber\\
                               \cos\theta J_{z}-\sin\theta J_{x} & 0 \nonumber\\
                             \end{array}
                           \right),\nonumber\\
\nonumber
\end{eqnarray}
where $I_{4}$ is a $4\times 4$ diagonal unit matrix and $J_{x}, J_{y}$ and $J_{z}$ are the angular momentum
matrices for a spin $3/2$ particle in the representation where $J_{z}$ is diagonal, i.e.
\begin{align}
I_4 &= \begin{pmatrix}
    1 & 0 & 0 & 0\\
    0 & 1 & 0 & 0\\
    0 & 0 & 1 & 0\\
    0 & 0 & 0 & 1\\
\end{pmatrix},
&
J_x &= \frac{\hbar}{2}\begin{pmatrix}
    0 & \sqrt{3} & 0 & 0\\
    \sqrt{3} & 0 & 2 & 0\\
    0 & 2 & 0 & \sqrt{3}\\
    0 & 0 & \sqrt{3} & 0\\
\end{pmatrix},
\end{align}
\begin{align}
J_y &= \frac{i\hbar}{2}\begin{pmatrix}
    0 & -\sqrt{3} & 0 & 0\\
    \sqrt{3} & 0 & -2 & 0\\
    0 & 2 & 0 & -\sqrt{3}\\
    0 & 0 & \sqrt{3} & 0\\
\end{pmatrix},
&
J_z &= \frac{\hbar}{2}\begin{pmatrix}
    3 & 0 & 0 & 0\\
    0 & 1 & 0 & 0\\
    0 & 0 & -1 & 0\\
    0 & 0 & 0 & -3\\
\end{pmatrix}.
\end{align}
Finally, using the Pauli spin matrices,
\begin{align}
\sigma_x &= \begin{pmatrix}
    0 & 1 \\
    1 & 0 \\
\end{pmatrix},
&
\sigma_y &= \begin{pmatrix}
    0 & -i \\
    i & 0 \\
\end{pmatrix},
&
\sigma_z &= \begin{pmatrix}
    1 & 0 \\
    0 & -1 \\
\end{pmatrix},
\end{align}
and the unit matrix in two dimensions, $I_{2},$
we write Eq.~(\ref{eq:HMDP1}) as
\begin{eqnarray}
\label{eq:HMDP2}
H_{M}&=&-\left(\frac{\hbar\Delta}{2}\right)\sigma_{z} \otimes I_{4}-\left(\frac{4\mu_{B}B}{5\hbar}\right)I_{2}\otimes J_{z}\nonumber\\
&&                           +\left(\frac{2\mu_{e}E}{5\hbar}\right)\sigma_{x}\otimes (J_{z}\cos\theta-J_{x}\sin\theta).\\
\nonumber
\end{eqnarray}

We note that every term in $H_{M}$ is a Kronecker product of two operators, denoted by $\otimes$. The first
operator in each product ($\sigma_{z}, I_{2}$, or $\sigma_{x})$ acts on a two level system, while the second
operator $(I_{4}, J_{z}$, or $J_{x})$ acts on a four level system. This form of the Hamiltonian reveals the
effective physics of the system:

The first term in the Hamiltonian simply corresponds to the lambda-doublet of transition frequency $\Delta$.
In the absence of any external fields, this doublet forms the two level system.

The second term appears in the presence of a magnetic field, $B$. Since parity is preserved by magnetic interactions,
$B$ does not mix the two lambda-doublet terms, leading to the presence of $I_{2}$. Within each lambda-doublet manifold,
however, the magnetic field removes the fourfold Zeeman degeneracy as indicated by the $J_z$. The two
\textit{doublet manifolds} now constitute the two level system.
This two level system no longer has a unique transition frequency, since each lambda-doublet manifold contains
four Zeeman states; however, it can still be manipulated coherently just like a standard two
level system, by using radiation at multiple frequencies. In this way, for example, exchanging the
population between the negative and positive energy doublet manifolds would implement the transformation
$\sigma_{z}\rightarrow -\sigma_{z}$, implying a $\pi$ rotation about the $x$ axis in pseudo-spin
space, as discussed below. If the relative populations and coherences within each lambda-doublet manifold are not
disturbed in the process, this transformation will not affect the dynamics of the angular momentum $J$.

The third term in Eq.~(\ref{eq:HMDP2}) is due to a non-zero electric field. Since the electric interaction does not
conserve parity, it mixes the lambda-doublet manifolds, via the $\sigma_{x}$. Also, angular momentum
along the $z$ axis is no longer generally conserved, as indicated by the presence of the $J_{x}.$ However,
when the electric and magnetic fields are collinear $(\theta=0$ or $\pi)$, angular momentum along $z$ \textit{is}
conserved, as only $J_{z}$ survives in the Hamiltonian.

As explained above, a $\pi$ rotation about the $x$ axis changes the sign of $\sigma_{z}$ and can be written as
$e^{-i\pi\sigma_{x}/2}\sigma_{z}e^{i \pi \sigma_{x}/2}=-\sigma_{z}$ \cite{SakuraiBook}. Thus, we can find the anti-commutation
$\{e^{-i \pi \sigma_{x}/2},\sigma_{z}\}=0.$
In a similar manner, for the spin $3/2$, a physical rotation by $\pi$ about the $y$ axis anticommutes with $J_{z}$ as well as $J_{x}:
\{e^{-i \pi J_{y}/\hbar},J_{z}\}=\{e^{-i \pi J_{y}/\hbar},J_{x}\}=0.$ Using these relations and the rule for anticommutation between
two Kronecker products $M_{1}\otimes M_{2}$ and $M_{3}\otimes M_{4}$ \cite{SteebBook},
\begin{eqnarray}
&&\{M_{1}\otimes M_{2},M_{3}\otimes M_{4}\}= \\
&&\frac{1}{2}\left([M_{1},M_{3}]\otimes [M_{2},M_{4}]+\{M_{1},M_{3}\}\otimes \{M_{2},M_{4}\}\right),\nonumber\\
\nonumber
\end{eqnarray}
it is straightforward to see that the rotation operator
\begin{equation}
C=e^{-i\pi\sigma_{x}/2}\otimes e^{-i\pi J_{y}/\hbar},
\end{equation}
anti-commutes with $H_{M}$ of Eq.~(\ref{eq:HMDP2}), i.e.
\begin{equation}
\label{eq:ACU}
H_{M}C+C H_{M}=0.
\end{equation}
If we now consider an eigenvector $\psi_{+}$ of $H_{M}$ with an eigenvalue $\lambda$, we can write
\begin{equation}
\label{eq:Eig1}
H_{M}\psi_{+}=\lambda\psi_{+}.
\end{equation}
Multiplying from the left by $C$ and using the anticommutation of Eq.~(\ref{eq:ACU}), we find the left hand
side of Eq.~(\ref{eq:Eig1}) reads $C H_{M}\psi_{+}=-H_{M}C\psi_{+},$ while the right hand side reads
$C(\lambda\psi_{+})=\lambda (C\psi_{+}).$ Equating the two sides, we arrive at
\begin{equation}
\label{eq:minus1}
H_{M}(C\psi_{+})=-\lambda(C\psi_{+}).
\end{equation}
This implies that $\psi_{-}=C\psi_{+}$ is an eigenfunction of $H_{M}$ with an eigenvalue $-\lambda$. We have thus
established that existence of the unitary operator $C$ leads to the $\pm \lambda$ pairing of eigenvalues in
the OH energy spectrum. The matrix representations of the operators $e^{-i\pi\sigma_{x}/2}$ and $e^{-i\pi J_{y}/\hbar}$ can be easily obtained; this leads to the following matrix for the operator $C$:
\begin{equation}
\label{eq:Cmat}
C=i\left(
  \begin{array}{cccccccc}
    0 & 0 & 0 & 0 & 0 & 0 & 0 & 1 \\
    0 & 0 & 0 & 0 & 0 & 0 & -1 & 0 \\
    0 & 0 & 0 & 0 & 0 & 1 & 0 & 0 \\
    0 & 0 & 0 & 0 & -1 & 0 & 0 & 0 \\
    0 & 0 & 0 & 1 & 0 & 0 & 0 & 0 \\
    0 & 0 & -1 & 0 & 0 & 0 & 0 & 0 \\
    0 & 1 & 0 & 0 & 0 & 0 & 0 & 0 \\
    -1 & 0 & 0 & 0 & 0 & 0 & 0 & 0 \\
  \end{array}
\right).
\end{equation}
That $C$ anticommutes with $H_{M}$ can therefore also be verified using simple matrix multiplication and Eqs.~(\ref{eq:Hmatrix})
and (\ref{eq:Cmat}). Clearly, the determinant $|C|=1\neq 0,$ and therefore $C^{-1}$ exists. Thus, the anticommutation relation
of Eq.~(\ref{eq:ACU}) may be written as $C^{-1}H_{M}C=-H_{M},$ implying that the result of the rotation $C$ is to simply invert
the sign of the Hamiltonian $H_{M}$. Such symmetries are often called \textit{chiral} and seem to have been noticed in only a
handful of physical systems \cite{McIntosh1962}. Perhaps the best known example is that of a free Dirac particle, where the
reflection symmetry of the spectrum follows from the anticommutation of the charge conjugation operator with the corresponding
Hamiltonian \cite{SakuraiBook}.

Interestingly, the structure of Eq.~(\ref{eq:HMDP2}) is that of a spin $1/2$ interacting with a spin $3/2$ system. Ising systems
with such spin combinations are of interest in condensed matter physics, as suitable models of ferrimagnetism \cite{Buendia1999,Weng1996}.
In principle, in such systems, an investigation of phenomena such as phase transitions requires a large number of interacting spins
in order to approximate well the thermodynamic limit. In practice, valuable information can be gained
from even two-spin ``Ising magnets" as demonstrated by the simulation of such a system by two trapped ions \cite{Porras2008}.
Our formulation of the Hamiltonian [see Eq.(\ref{eq:HMDP2})] suggests that the OH molecule could be used to simulate an
imbalanced (i.e. unequal spin) Ising magnet. By changing the angle $\theta$ between the magnetic and electric fields, the interactions
may be continuously varied from purely transverse $(\sim \sigma_{x}J_{x})$ to transverse-longitudinal $(\sim \sigma_{x}J_{z})$
\cite{Buendia1999}. Full quantum tomography of similar ground state manifolds has already been experimentally demonstrated in
atomic systems \cite{Smith2013}. A similar experiment with OH could provide complete knowledge of the
$X^{2}\mathrm{\Pi}_{3/2}$ density matrix, including information about either subsystem (doublet or rotor, obtained by tracing
over the appropriate state space), as well as correlations between the two.

In conclusion, we have solved exactly the effective Hamiltonian of the OH $X^{2}\Pi_{3/2}$ state
molecule in combined electric and magnetic fields, neglecting hyperfine structure. We have identified explicitly the source of the
reflection symmetry in the spectrum that makes the problem exactly soluble. Our analysis opens the way to a more precise
and insightful characterization of the magneto-electrostatic manipulation of OH, and to its use for quantum simulation.
We thank B. Sawyer, B. Stuhl, M. Hummon, M. Yeo, S. Rajeev, A. Das and M. Lahiri for useful discussions and
J. L. Bohn and G. Quemener for correspondence regarding their recent work \cite{Bohn2013}. Z.H. would like to thank RIT for
a summer research award. M. K. is grateful for support through the National Science Foundation (award No. 1068112).


\begin{thebibliography}{27}
\expandafter\ifx\csname natexlab\endcsname\relax\def\natexlab#1{#1}\fi
\expandafter\ifx\csname bibnamefont\endcsname\relax
  \def\bibnamefont#1{#1}\fi
\expandafter\ifx\csname bibfnamefont\endcsname\relax
  \def\bibfnamefont#1{#1}\fi
\expandafter\ifx\csname citenamefont\endcsname\relax
  \def\citenamefont#1{#1}\fi
\expandafter\ifx\csname url\endcsname\relax
  \def\url#1{\texttt{#1}}\fi
\expandafter\ifx\csname urlprefix\endcsname\relax\def\urlprefix{URL }\fi
\providecommand{\bibinfo}[2]{#2}
\providecommand{\eprint}[2][]{\url{#2}}

\bibitem[{avd()}]{Ticknor2005}
\bibinfo{note}{C. Ticknor and J. L. Bohn, Phys. Rev. A {\bf 71}, 022709 (2005).}

\bibitem[{avd()}]{Avdeenkov2003}
\bibinfo{note}{A. V. Avdeenkov and J. L. Bohn, Phys. Rev. Lett. {\bf 90}, 043006 (2003).}

\bibitem[{saw()}]{Sawyer2008}
\bibinfo{note}{B. C. Sawyer, B. K. Stuhl, D. Wang, M. Yeo and J. Ye, Phys. Rev. Lett. {\bf 101}, 203203 (2008).}

\bibitem[{tsc()}]{Tscherbul2010}
\bibinfo{note}{T. V. Tscherbul, Z. Pavlovic, H. R. Sadeghpour, R. Cote, and A. Dalgarno,
Phys. Rev. A {\bf 82}, 022704 (2010).}

\bibitem[{hud()}]{Hudson2006}
\bibinfo{note}{E. R. Hudson, H. J. Lewandowski, B. C. Sawyer and J. Ye, Phys. Rev. Lett. {\bf 96}, 143004 (2006).}

\bibitem[{koz()}]{Kozlov2009}
\bibinfo{note}{M. G. Kozlov, Phys. Rev. A {\bf 80}, 022118 (2009).}

\bibitem[{lev()}]{Lev2006}
\bibinfo{note}{B. L. Lev, E. R. Meyer, E. R. Hudson, B. C. Sawyer, J. L. Bohn and J. Ye , Phys. Rev. A
{\bf 74}, 061402(R) (2006).}

\bibitem[{que()}]{Quemener2012}
\bibinfo{note}{B. K. Stuhl, M. T. Hummon, M. Yeo, G. Quemener, J. L. Bohn and J. Ye, Nature {\textbf 492}, 396 (2012).}

\bibitem[{mei()}]{Meijer2009}
\bibinfo{note}{S. A. Meek, H. Conrad and G. Meijer , Science {\textbf 324}, 1699 (2009).}

\bibitem[{boc()}]{Bochinski2004}
\bibinfo{note}{J. R. Bochinski. E. R. Hudson, H. J. Lewandowski, and J. Ye, Phys. Rev. A {\bf 70}, 043410 (2004).}

\bibitem[{mee()}]{Meerakker2005}
\bibinfo{note}{S. Y. T. van de Meerakker, P. H. M. Smeets, N. Vanhaecke, R. T. Jongma, and G. Meijer,
Phys. Rev. Lett. {\bf 94}, 023004 (2005).}

\bibitem[{mee()}]{Sawyer2007}
\bibinfo{note}{B. C. Sawyer, B. L. Lev, E. R. Hudson, B. K. Stuhl, M. Lara, J. L. Bohn, and J. Ye,
Phys. Rev. Lett. {\bf 98}, 253002 (2007).}

\bibitem[{stu()}]{Stuhl2012}
\bibinfo{note}{B. K. Stuhl, M. Yeo, B. C. Sawyer, M. T. Hummon and J. Ye, Phys. Rev. A {\bf 85}, 033427 (2012).}

\bibitem[{sch()}]{Schmelcher2013}
\bibinfo{note}{M. Garttner, J. J. Olmiste, P. Schmelcher, and R. Gonzalez-Ferez, arxiv:1301.4586v1 (2013).}

\bibitem[{boh()}]{Bohn2013}
\bibinfo{note}{J. L. Bohn and G. Quemener, arxiv:1301.2590v1 (2013).}

\bibitem[{bue()}]{Buendia1999}
\bibinfo{note}{G. M. Buendia and R. Cardona, Phys. Rev. B {\bf 59}, 6784 (1999).}

\bibitem[{lar()}]{Lara2008}
\bibinfo{note}{M. Lara, B. L. Lev and J. L. Bohn, Phys. Rev. A {\bf 78}, 033433 (2008).}

\bibitem[{way()}]{Wayne1990}
\bibinfo{note}{R. P. Wayne, Sci. Prog. {\bf 74}, 379 (1990).}

\bibitem[{war()}]{Wardle2002}
\bibinfo{note}{M. Wardle and F. Yusuf-Zadeh, Science {\bf 296}, 2350 (2002).}

\bibitem[{smi()}]{Smith2002}
\bibinfo{note}{G. P. Smith, J. Luque, C. Park, J. B. Jeffries, and D. R. Crosley, Combust. Flame {\bf 131}, 59 (2002).}

\bibitem{SchmelcherBook} P. Schmelcher and W. Schweizer, \textit{Atoms and molecules in strong external fields},
(Springer, Berlin, 1998).

\bibitem[{mer()}]{Merriman1892}
\bibinfo{note}{M. Merriman, Bull. New York Math. Soc., {\textbf 1}, 202 (1892).}

\bibitem[{\citenamefont{Sakurai}(1967)}]{SakuraiBook}
\bibinfo{author}{\bibfnamefont{J. J.}~\bibnamefont{Sakurai}},
\emph{\bibinfo{title}{Advanced Quantum Mechanics}}
  (\bibinfo{publisher}{Addison-Wesley},
\bibinfo{address}{United States},
\bibinfo{year}{1967}).

\bibitem{SteebBook} W. H. Steeb and Y. Hardy, \textit{Matrix Calculus and Kronecker Product: A Practical Approach to Linear and Multilinear Algebra},
(World Scientific, United States, 2011).

\bibitem[{mci()}]{McIntosh1962}
\bibinfo{note}{H. V. McIntosh, J. Mol. Spec. {\bf 8}, 169 (1962).}

\bibitem[{wen()}]{Weng1996}
\bibinfo{note}{X. M. Weng and Z. Y. Li, Phys. Rev. B {\bf 53}, 12142 (1996).}

\bibitem[{por()}]{Porras2008}
\bibinfo{note}{A. Friedenauer, H. Schmitz, J. T. Glueckert, D. Porras and T. Schaetz, Nature Physics {\bf 4}, 757 (2008).}

\bibitem[{smi()}]{Smith2013}
\bibinfo{note}{A. Smith, C. A. Riofrio, B. E. Anderson, H. Sosa-Martinez, I. H. Deutsch and P. S. Jessen, Phys. Rev. A {\bf 87},
030102(R)(2013).}
\end{thebibliography}
\end{document}